\documentstyle[twoside,fleqn,epsfig]{article}

\def\be{\begin{equation}}
\def\ee{\end{equation}}
\def\bea{\begin{eqnarray}}
\def\eea{\end{eqnarray}}
\newcommand{\lsim}{\mathrel{\mathop{\kern 0pt \rlap
  {\raise.2ex\hbox{$<$}}}
  \lower.9ex\hbox{\kern-.190em $\sim$}}}
\newcommand{\gsim}{\mathrel{\mathop{\kern 0pt \rlap
  {\raise.2ex\hbox{$>$}}}
  \lower.9ex\hbox{\kern-.190em $\sim$}}}

\newcommand{\AmS}{{\protect\the\textfont2
  A\kern-.1667em\lower.5ex\hbox{M}\kern-.125emS}}
\normalsize
\begin{document}
\Large
\begin{flushright}
{\bf ROM2F/2007/15 \\} 
{\bf September 2007 \\}
{\bf submitted for publication \\}
\end{flushright}
\normalsize

\baselineskip=0.65cm
\vspace*{0.5cm}

\begin{center}
\Large \bf
Possible implications of the channeling effect in NaI(Tl) crystals \\
\rm
\end{center}

\vspace{0.5cm}
\normalsize

\noindent \rm R.\,Bernabei,~P.\,Belli,~F.\,Montecchia,~F.\,Nozzoli

\noindent {\it Dip. di Fisica, Universit\`a di Roma ``Tor Vergata"
and INFN, sez. Roma ``Tor Vergata", I-00133 Rome, Italy}  

\vspace{3mm}

\noindent \rm F.\,Cappella, A.\,Incicchitti,~D.\,Prosperi

\noindent {\it Dip. di Fisica, Universit\`a di Roma ``La Sapienza"
and INFN, sez. Roma, I-00185 Rome, Italy}

\vspace{3mm}

\noindent \rm R.\,Cerulli

\noindent {\it Laboratori Nazionali del Gran Sasso, INFN, Assergi, Italy}

\vspace{3mm}

\noindent \rm C.J.\,Dai,~H.L.\,He,~H.H.\,Kuang,~J.M.\,Ma,~X.H.\, Ma, ~X.D.\,Sheng,~Z.P.\,Ye\footnote{also:
University of Jing Gangshan, Jiangxi, China},R.G.\,Wang, Y.J.\, Zhang

\noindent {\it IHEP, Chinese Academy, P.O. Box 918/3, Beijing 100039, China}

\vspace{0.5cm}
\normalsize

\begin{abstract}

The channeling effect of low energy ions along the crystallographic axes and planes 
of NaI(Tl) crystals is discussed in the framework of corollary investigations on  WIMP
Dark Matter candidates. In fact, the modeling of this existing effect implies a more complex  
evaluation of the luminosity yield for low energy recoiling Na and I ions. 
In the present paper related phenomenological arguments are developed and  
possible implications are discussed at some extent. 

\end{abstract}

{\it Keywords:} Dark Matter; WIMP; underground 
Physics

{\it PACS numbers:} 95.35.+d

\section{Introduction}

\begin{figure} [!t]
\centering
\includegraphics[width=9.0cm] {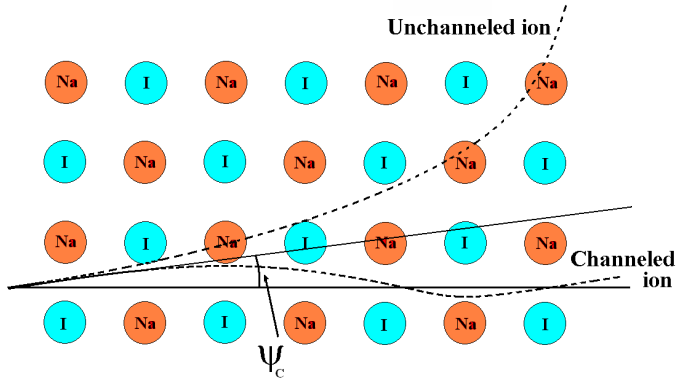}
\caption{Simplified schema of the channeling effect in the NaI(Tl) lattice. The axial channeling
occurs when the angle of the motion direction of an ion with the respect to the crystallographic axis
is less than a characteristic angle, $\Psi_c$, depicted there (see for details Sec. 2). Two examples for 
channeled and unchanneled ions are also shown (dashed lines).}
\label{fg:schema_chan}
\end{figure}

It is known that ions (and, thus, also recoiling nuclei) move in a crystal in a different
way than in amorphous materials. In particular, ions moving (quasi-) parallel to crystallographic
axes or planes feel the so-called ``channeling effect'' and show an anomalous deep penetration into the
lattice of the crystal \cite{chan,nel63,oth2}; see Fig. \ref{fg:schema_chan}.

For example, already on 1957, a penetration of $^{134}$Cs$^+$ ions into a Ge crystal 
was observed to a depth of about 1000 \AA \, \cite{Bre56}, larger than that expected in the case
the ions would cross amorphous Ge ($\simeq 50$ \AA). 
Afterwards, high intensities of H$^+$ ions at 75 keV transmitted through thick (3000-4000 \AA)
single-crystal gold films in the $<110>$ directions were detected \cite{nel63}.
Other examples for keV range ions have been shown in ref. \cite{Ras05} where 3 keV P$^+$ ions moving into layers of  
500 \AA \, of various crystals were studied. 

The channeling effect is also exploited in high energy Physics e.g.
to extract high energy ions from a beam by means of bent crystals or
to study diffractive Physics by analysing scattered ions along the beam direction (see e.g. ref.
\cite{Baur00}).

Recently \cite{droby2} it has been pointed out the possible role which this effect can play 
in the evaluation of the detected energy of recoiling nuclei in crystals, 
such as the NaI(Tl)\footnote{For completeness, it is worth to note that luminescent 
response for channeling in NaI(Tl) was already studied in ref. \cite{lunt} for MeV-range ions.}.

In fact, the channeling effect can occur in crystalline materials due to correlated collisions of ions with target atoms.
In particular, the ions through the open channels have ranges much larger than 
the maximum range they would have if their motion would be either in other directions 
or in amorphous materials. Moreover, 
when a low-energy ion goes into a channel, its energy losses are mainly 
due to the electronic contributions. This implies that a channeled ion transfers 
its energy mainly to electrons rather than to the nuclei in the lattice and, thus,  
its quenching factor (namely the ratio between
the detected energy in keV electron equivalent [keVee] 
and the kinetic energy of the recoiling nucleus in keV) approaches the unity.  

It is worth to note that this fact can have a role in corollary analyses 
in the Dark Matter particle direct detection experiments,
when WIMP (or WIMP-like) candidates are considered. In fact, since
the routine calibrations of the detectors are usually performed by using  
$\gamma$ sources (in order to avoid induced radioactivity in the materials), 
the quenching factor is a key quantity to derive the energy of the recoiling nucleus after an elastic scattering.
Generally, for scintillation and ionization detectors this factor has been inferred so far 
by inducing tagged recoil nuclei through neutron elastic scatterings \cite{Mis_neut}; however, as it will be discussed
in Sec. 3, the usual analysis carried out on similar measurements does not allow to account for the channeled events.
A list of similar values for various nuclei in different 
detectors can be found e.g. in ref. \cite{RNC}.
In particular, commonly in the interpretation of the dark matter direct detection 
results in terms of WIMP (or WIMP-like) candidates the quenching factors are assumed to be 
constant values without considering e.g. their energy dependence, the properties 
of each specific used detector and the experimental uncertainties. 
An exception was in the DAMA/NaI corollary model dependent analyses 
for WIMP (or WIMP-like) candidates \cite{RNC,ijmd,epj06,ijma2} where  
at least some of the existing uncertainties on the $q_{Na}$ and $q_I $ values, 
measured with neutrons, were included.

In this paper the possible impact of the channeling effect in NaI(Tl) crystals is discussed in a
phenomenological framework and comparisons on some of the corollary analyses 
carried out in terms of WIMP (or WIMP-like) candidates \cite{RNC,ijmd,epj06,ijma2},
on the basis of the 6.3 $\sigma$ C.L. DAMA/NaI model independent evidence for
particle Dark Matter in the galactic halo\footnote{We remind that 
various possibilities for some of the many possible astrophysical, nuclear and particle Physics scenarios have 
have been analysed by DAMA itself both for some WIMP/WIMP-like candidates and for light bosons \cite{RNC,ijmd,epj06,ijma2,ijma}, 
while other corollary analyses are also available in literature, such as e.g.
refs. \cite{Bo03,Bo04,Botdm,khlopov,Wei01,foot,Saib,droby1,droby2}. 
Many other scenarios can be considered as well.}, are given. 

\section{Luminosity, range and channeling in NaI(Tl) crystals}

The stopping power of an ion inside an amorphous material
is given by the sum of two effects: its interaction with the nuclei ($n$) of the material 
and its interaction with the binding electrons ($e$)\cite{Lind1}. 

If $E$ is the energy of the ion at any point $x$
along the path, its stopping power can be written as: 
\begin{equation}
\frac{dE_{ion}}{dx}(E) = \frac{dE_{ion-n}}{dx}(E) + \frac{dE_{ion-e}}{dx}(E)
\label{eq:1}
\end{equation}
where $\frac{dE_{ion-n}}{dx}$ and $\frac{dE_{ion-e}}{dx}(E)$ are the nuclear and the electronic 
stopping powers of the ion, respectively. They can be evaluated 
-- following the theory firstly developed in ref. \cite{Lind1} -- 
by using some available packages, as the SRIM code \cite{SRIM}.

We have to stress that the detectable light produced by a charged particle (either electron or ion)
in scintillator detectors mostly arises from the energy loss in the electronic interactions.
Thus, the differential luminosities in scintillators, $\frac{dL_e}{dx}$ and $\frac{dL_{ion}}{dx}$
for electrons and ions, respectively, 
can be written as:
\begin{eqnarray}
\frac{dL_{e}}{dx} &=& \alpha \frac{dE_{e-e}}{dx} \mbox{\hspace{5.5cm} for electrons} \nonumber \\
\frac{dL_{ion}}{dx} &=& \alpha \frac{dE_{ion-e}}{dx}(E) = \alpha \times \frac{dE_{ion}}{dx} \times q'(E) 
\mbox{\hspace{1.6cm} for ions (recoils)}
\label{eq:2}
\end{eqnarray}
where $\alpha$ is a proportionality constant, $\frac{dE_{e-e}}{dx}$ is the stopping power of 
an electron in the material and
$q'(E) = \left( \frac{dE_{ion-e}}{dx}(E)\right) /  \left( \frac{dE_{ion-n}}{dx}(E) + \frac{dE_{ion-e}}{dx}(E) \right)$ 
is defined as ``differential quenching factor''. 

The total detected luminosities, $L=\int_{path}\frac{dL}{dx} dx$, for electrons and ions can be written in the form: 
\begin{eqnarray}
L_{e} &=& \alpha \int_{path} \frac{dE_{e-e}}{dx} dx = \alpha E_{e} \mbox{\hspace{3.8cm} for electrons} \nonumber \\
L_{ion} &=& \alpha \int_{path} q'(E) \times \frac{dE_{ion}}{dx} dx 
\equiv \alpha \times q_1(E_{ion}) \times E_{ion} \mbox{\hspace{0.2cm} for ions (recoils) \label{l_ion};}  
\end{eqnarray}
in addition, the range of an ion is:
\begin{equation}
{\textsf R}_{ion}(E) = \int_0^E \frac{dE'}{dE_{ion}/dx} \equiv q_2(E) \times \int_0^E \frac{dE'}{dE_{ion-e}/dx} 
= q_2(E) \times {\textsf R}_{e}(E).
\label{ran_ion}
\end{equation}
In eq. (\ref{l_ion}) and (\ref{ran_ion}) $q_1(E) = \frac{1}{E} \int_0^E q'(E')dE'$ is the ``light quenching factor''
and $q_2(E) = \int_0^E \frac{dE'}{dE_{ion}/dx}/\int_0^E \frac{dE'}{dE_{ion-e}/dx}$ is the 
``range quenching factor''.
In the energy region of interest for the dark matter detection $q_1(E) \simeq q_2(E)$ 
within 10-20\%. 
In particular, the values of the quenching factors for recoiling nuclei in detectors made of amorphous materials
are well below the unity in the keV energy region. 

The situation changes 
when the detector is either a crystal or a multi-crystalline material (the size of 
a single crystal has to be larger than few thousands of \AA). In this case
the luminosity depends on whether the recoiling nucleus is (quasi-) parallel 
to the crystallographic axes or planes or not. In the first case, since the  
energy losses of the ion are mainly due to the electronic contributions, 
the penetration (and the range) of the ion becomes much larger, of the order of
${\textsf R}_{e}$, and the
quenching factor approaches the unity.

The theory of ion channeling in crystals has been developed e.g. in ref. \cite{Lind2,oth2}.
In particular, this theory deals with channeling of low energy, high mass ions as a separate 
case from high energy, low mass ions.
Here, we only remind that the condition for a low energy ion and a recoiling nucleus to be  
axially channeled along a certain string of atoms in the lattice is linked to 
a critical angle, $\Psi_c$ (see Fig. \ref{fg:schema_chan}); for details refer to \cite{Lind2,oth2}. 
When the ion (recoiling nucleus) has a moving direction
with an angle $\Psi$ with respect to this string lower than  $\Psi_c$, it is axially channeled.

The critical angles for axial channeling is given by \cite{Lind2,oth2}:
\begin{equation}
\Psi_c = \sqrt{\frac{Ca_{TF}}{d\sqrt{2}} \Psi_1 }
\end{equation}
\noindent where $C^2 \simeq 3$ is the Lindhard's constant and $d$ is the inter-atomic spacing 
in the crystal along the channeling direction. 
The Thomas-Fermi radius, $a_{TF}$, can be written as \cite{Lind2,oth2}:
\begin{equation}
a_{TF} = \frac{0.8853 a_0}{ \left(\sqrt{Z_1} + \sqrt{Z_2} \right)^{2/3} }
\end{equation}
where $Z_1$ and $Z_2$ are the atomic numbers of the projectile (recoiling nucleus) and target atoms, respectively;
$a_0 = 0.529 \AA$ is the Bohr radius.

The characteristic angle $\Psi_1$ is defined as a function of the ion (recoiling nucleus) energy, $E$, by:
\begin{equation}
\Psi_1 = \sqrt{\frac{2Z_1Z_2e^2}{Ed}}
\end{equation}
where $e$ is the electron's charge. The condition $\Psi<\Psi_c$ 
for axial channeling is valid for $\Psi_1>\Psi_{1,lim} = \frac{a_{TF}}{d}$, that is
for $E<E_{lim}=\frac{2Z_1Z_2e^2 d}{a_{TF}^2} $ \cite{Lind2,oth2}. Hence, typical values for NaI(Tl) crystal
assure that for recoil's energies less than 170 keV the quoted formulas hold. 
For completeness, we just remind that it has also been suggested that the critical 
angles may slightly depend on the temperature \cite{Mor71}.
Moreover, the critical angles at low energy have a weaker dependence on ion energy than those 
at higher energy. In fact, at higher energy, the critical angle is $\simeq C\Psi_1$ \cite{Lind2,oth2}. 

\vspace{0.3cm}

In the case of planar channeling for low energy ions the critical angle can be written as
\cite{Guha} (see also ref. \cite{Lind2}):
\begin{equation}
\theta_{pl} = a_{TF} \sqrt{Nd_p} \left( \frac{Z_1Z_2e^2}{Ea_{TF}} \right)^{1/3} ,
\end{equation}
where $N$ is the atomic number density and $d_p$ is the inter-plane spacing.
The dependence of $\theta_{pl}$ on the energy is weaker than that  
at higher energy, where it can be written as \cite{Baur00,Pic69}:
\begin{equation}
\theta_{pl} = a_{TF} \sqrt{Nd_p} \left( \frac{2Z_1Z_2e^2C}{Ea_{TF}} \right)^{1/2}.
\end{equation}

\begin{figure} [ht]
\centering
\vspace{-0.8cm}
\includegraphics[width=7.0cm] {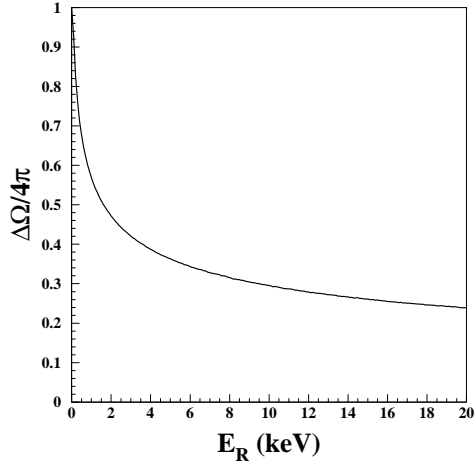}
\vspace{-0.6cm}
\caption{Fraction of solid angle interested by both axial and planar channeling
in NaI(Tl) crystals as a function of the energy of the recoiling nuclei, calculated
according to the text.
In these calculations just the lower index crystallographic axes and planes have been considered:
for the axial channeling: $<100>$, $<110>$, $<111>$ and for the planar channeling: 
$\{100\}$, $\{110\}$, $\{111\}$.}
\label{fg:solid_ang}
\vspace{-0.4cm}
\end{figure}

Taking into account the critical angles for axial and planar channeling
in NaI(Tl), we have calculated by Monte Carlo method 
the solid angle interested by both axial and planar channeling in NaI(Tl) crystals 
as a function of the energy of the recoiling nuclei, $E_R$; see Fig. \ref{fg:solid_ang}.
Moreover, just the lower index crystallographic axes and planes have been considered,
for the axial channeling: $<100>$, $<110>$, $<111>$ and for the planar channeling: 
$\{100\}$, $\{110\}$, $\{111\}$. 

In this way, the estimated light response of a NaI(Tl) crystal scintillator to Sodium and Iodine recoils 
at given energy has been studied taking into account the channeling effect in the considered modeling.
For a given nucleus $A$ with recoil energy $E_R$ the response of a NaI(Tl) crystal scintillator
can be written as $\frac{dN_A}{dE_{det}}(E_{det}|E_R)$, where $E_{det}$ is the detected energy.
By the definition: $\int_0^\infty \frac{dN_A}{dE_{det}}(E_{det}|E_R) dE_{det} = 1$.
In most cases of the Dark Matter direct detection field -- that is without including the channeling effect --
the light response is assumed equal to a Dirac delta function: 
$\frac{dN_A}{dE_{det}}(E_{det}|E_R) = \delta(E_{det} - q_AE_R) $, where 
$q_A$ is the value (assumed constant) of the quenching factor of the unchanneled $A$ nucleus recoils. 

\begin{figure} [!ht]
\centering
\vspace{-1.2cm}
\includegraphics[width=9.5cm] {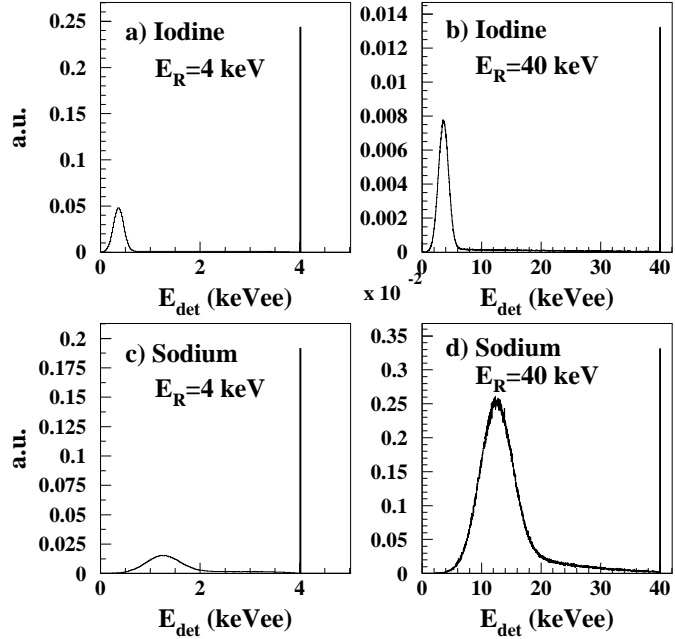}
\vspace{-0.4cm}
\caption{Examples of light responses in terms of keVee, $\frac{dN_A}{dE_{det}}(E_{det}|E_R)$, 
for Iodine recoils of 4 keV ($a$)
and of 40 keV ($b$) and for Sodium recoils of 4 keV ($c$) and of 40 keV ($d$)
in the modeling given in the text.
In this calculation the quenching factors for Sodium and Iodine recoils 
in amorphous or out of channel NaI(Tl) are assumed at the mean values 
given in ref. \cite{Psd96}. Just to emphasize the effect of the channeling, the broadening
due to the energy resolution of the detector has not been included here. 
The peaks corresponding to fully channeled events ($q\sim1$) and to fully quenched events (broadened 
by the straggling) are well evident; in the middle 
events, which have been de-channeled at least once, are also visible.
It is possible to note that e.g. in the case of Iodine recoils the fully channeled events are 
about 25\% at 4 keV; this percentage becomes smaller, about 1\% at 40 keV.}
\label{fg:response}
\end{figure}

The evaluation of $\frac{dN_A}{dE_{det}}$, when accounting for channeling effect,
has been realised by means of a Monte Carlo code; the path of a recoiling nucleus, at a given recoil 
energy  $E_R$, has been calculated under the following reasonable and cautious assumptions: 

\newcounter{a}
\begin{list}{\roman{a}}{\usecounter{a}}
\item ) isotropic distribution of the recoils;
\item ) in the case the recoil would enter in a channel, a de-channeling can occur due to some
interactions with impurities in the lattice, as Tl luminescent dopant centers. The 
probability density of such a process is assumed to be: $p(x) = \frac{1}{\lambda}e^{-x/\lambda}$, 
with $\lambda = 1200$ \AA, that is the average distance among the Tl centers;
\item ) the energy losses by the recoil nuclei in a channel just depend
      on the electronic stopping power (see eq. (\ref{eq:1}));
\item ) the energy losses by the recoil nuclei in a channel are converted into scintillation light
      with a quenching factor $\sim 1$;
\item ) if a recoil is de-channeled due to a nuclear interaction, it can either re-enter into another channel 
      or not. The differential distribution of the nuclear interaction is assumed to be isotropic;
\item ) in case the recoil would not enter into a channel, it is cautiously assumed that it 
      stops and its released energy is converted into scintillation light
      with the quenching factor of unchanneled events. In this case the straggling is considered as evaluated by the SRIM code \cite{SRIM}. 
\end{list}

In Fig. \ref{fg:response} few examples of light responses in terms of keVee for Iodine recoils of 4 keV ($a$)
and of 40 keV ($b$) and for Sodium recoils of 4 keV ($c$) and of 40 keV ($d$) are given.
In these calculations the quenching factors for Sodium and Iodine 
recoils in amorphous or out of channel NaI(Tl) are assumed at the 
mean values given in ref. \cite{Psd96}. Just to emphasise the effect of the channeling, the broadening
due to the energy resolution of the detector has not been included in this figure. 
The peaks corresponding to fully channeled events ($q\sim1$) 
and to fully quenched events (broadened by the straggling) are well evident;  in the middle 
events, which have been de-channeled at least once, are also visible.

\begin{figure} [ht]
\centering
\vspace{-0.8cm}
\includegraphics[width=7.0cm] {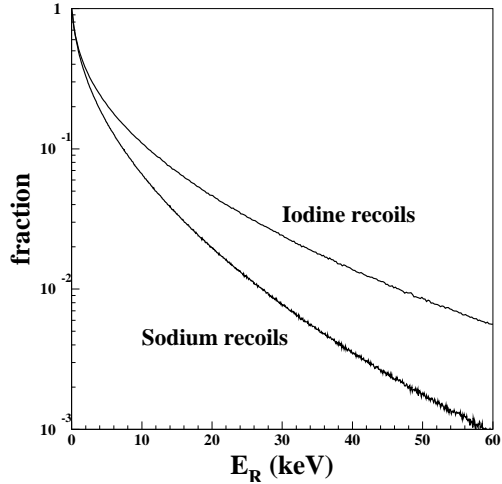}
\vspace{-0.4cm}
\caption{Fraction of events with quenching factor $\simeq 1$, that is fully channeled events,
as a function of the energy of the recoiling nuclei in NaI(Tl) crystals according to the modeling
described in the text.}
\label{fg:frac}
\end{figure}

Finally, in Fig. \ref{fg:response} it is possible to note 
that the number of fully channeled ($q\sim1$) events strongly decreases when increasing the recoil energy:
they are $\sim25\%$ at 4 keV and $\sim1\%$ at 40 keV for Iodine recoils and
$\sim18\%$ at 4 keV and $\sim0.3\%$ at 40 keV for Sodium recoils. These behaviours
are depicted in Fig. \ref{fg:frac}.

\section{Some comments}

Let us now analyse the phenomenologies 
connected both with the data on nuclear recoils induced by 
neutron scatterings and with the WIMP (or WIMP-like) direct detection in the light of the presence of 
the channeling effect.

\begin{figure} [ht]
\centering
\vspace{-0.8cm}
\includegraphics[width=6.0cm] {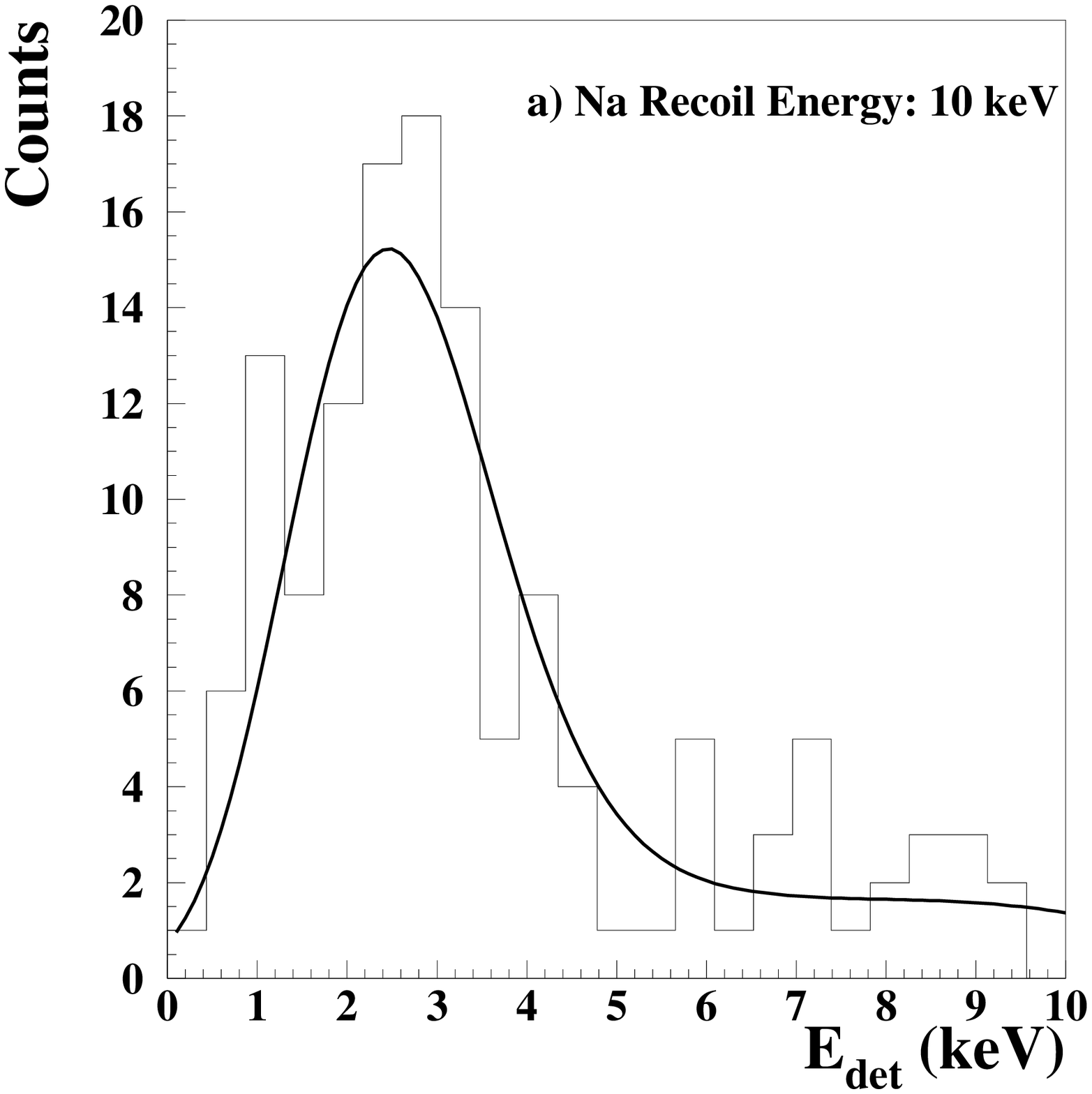}
\includegraphics[width=6.0cm] {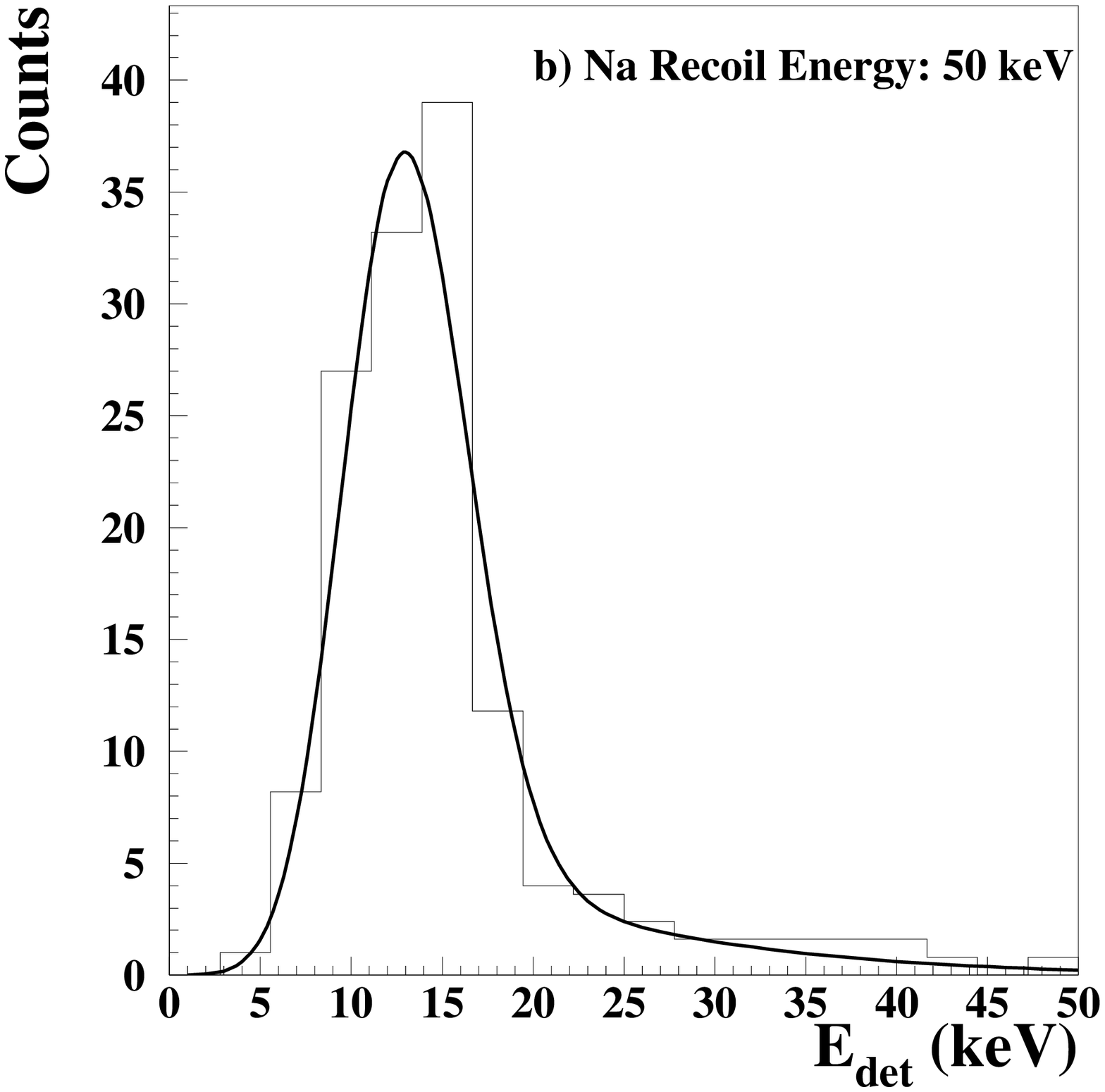}
\vspace{-0.2cm}
\caption{Examples of neutron calibrations of NaI(Tl) detectors at low recoil energy.
In particular, the energy responses of NaI(Tl) detectors to 
Sodium recoils of 10 keV (left panel) \cite{neut06} and of 50 keV (right panel) \cite{neut03}
are shown;  the peaks corresponding to the quenched events are well clear.
The superimposed continuous curves have been calculated as those of Fig. \ref{fg:response},
obviously broadening them by the energy resolution of the corresponding detector.
The fully channeled peaks ($q\sim1$), which contain in these cases only the 6\% and 0.15\% of the events respectively,
are smeared out by the energy resolution and can just contribute to the higher energy tails 
in the energy spectra.}
\label{fg:neut}
\end{figure}

In particular, Fig. \ref{fg:neut} shows some examples of neutron calibrations 
of NaI(Tl) detectors at relatively low recoil energy. 
There the energy responses of the used NaI(Tl) detectors to 
Sodium recoils of 10 keV \cite{neut06} and of 50 keV \cite{neut03}
are reported as solid histograms; the peaks corresponding to the quenched events are well clear.
The superimposed continuous curves have been calculated as those of Fig. \ref{fg:response},
obviously broadening them by the energy resolution of the corresponding detector.
The fully channeled peaks ($q\sim1$), which in these cases can 
contain only the 6\% and 0.15\% of the events respectively (see Fig. \ref{fg:frac}),
are smeared out by the energy resolution and only contribute to the higher energy tails 
in the energy spectra.

\noindent Thus, the simple analysis of Fig. \ref{fg:neut} shows that the neutron data can
contain channeled events,
but -- owing to the low-statistics of these measurements, to the small effect looked for 
and to the energy resolution -- 
they cannot easily be identified. Moreover, as already shown, 
the channeling effect becomes less important at increasing energy 
and gives more suppressed contributions in the neutron scattering data. For Iodine recoils the situation 
is even worse. Therefore, there is no hope to single out the channeling effect 
in the already-collected neutron data. 

\vspace{0.3cm}

On the other hand, the accounting of the channeling effect can give a significant impact 
in the sensitivities of the Dark Matter direct detection methods when WIMP (or WIMP-like)
candidates are considered. In particular in par.4 and 5 we will show that
lower cross sections are explorable in given models for WIMP and WIMP-like candidates 
by crystal scintillators, such as NaI(Tl) (up to more than a factor 10 in some mass range). 
Moreover:

\newcounter{b}
\begin{list}{\roman{b}}{\usecounter{b}}
\item ) similar situation holds for purely ionization detectors, as Ge (HD-Moscow - like experiments);
\item ) the loss of sensitivity occurs when pulse shape discrimination is used in crystal scintillators (KIMS); 
      in fact, the channeled events ($q \simeq 1$) are probably lost;
\item ) no enhancement can be present in liquid noble gas experiments (DAMA/LXe, WARP, XENON, ...);
\item ) no enhancement is possible for bolometer experiments; on the contrary some loss of sensitivity
      is expected since events (those with $q_{ion} \simeq 1$) 
      are lost by applying some discrimination procedures, based on $q_{ion} << 1$. 
\end{list}

\section{Application to the WIMP-nucleus elastic scattering}

Let us now consider the case of WIMP (or WIMP-like) elastic scattering
on target nuclei. In particular, the expected differential counting rate
of recoils induced by WIMP-nucleus elastic scatterings has to be evaluated 
in given astrophysical, nuclear and particle physics scenarios,
also requiring assumptions on all the parameters involved in the calculations and the 
proper consideration of the related uncertainties (for some discussions 
see e.g. \cite{RNC,ijmd,ijma,epj06,ijma2}).
Hence, the proper accounting for the channeling effects 
must be considered as an additional uncertainties
in the evaluation of the expected differential counting rate. 
The usual hypothesis that just one component of the dark halo can produce elastic 
scatterings on nuclei will be assumed here. In addition, the presence of the existing Migdal effect
and the possible SagDEG contribution -- we discussed in refs. \cite{ijma2} and \cite{epj06} respectively --
will be not included here for simplicity.
Thus, 
for every target specie $A$, the expected 
distribution of the detected energy can be written as
a convolution between the light response function, 
$\frac{dN_A}{dE_{det}}$, defined in the previous section, and
the differential distribution  
produced in the WIMP-nucleus elastic scattering:
\begin{equation}
\frac{dR^{(ch)}_A}{dE_{det}}(E_{det}) = \int
\frac{dN_A}{dE_{det}} (E_{det}|E_{R})
\frac{dR_A}{dE_{R}}(E_{R}) dE_R  \; . 
\label{eq:rt}
\end{equation}
The differential energy distribution of recoils, as function of the recoil energy $E_R$,
is:
\begin{equation}
\frac{dR_A}{dE_{R}}(E_{R}) = 
N_{T}\frac{\rho_{W}}{m_W}\int^{v_{max}}_{v_{min}(E_{R})}
\frac{d\sigma}{dE_{R}}(v,E_{R}) v f(v) dv  \; . 
\end{equation}
In this formula: 
i) $N_T$ is the number of target nuclei of A specie; 
ii) $\rho_W = \xi \rho_0$, where
$\rho_0$ is the local halo density and $\xi \leq 1$ 
is the fractional amount of local WIMP density;
iii) $m_W$ is the WIMP mass;
iv) $f(v)$ is the WIMP velocity ($v$) distribution in the Earth frame;
v) $v_{min} = \sqrt{\frac {m_A \cdot E_R}{2 m^2_{WA}}}$ 
($m_{A}$ and $m_{WA}$ are the nucleus mass and 
the reduced mass of the WIMP-nucleus system, respectively);
vi) $v_{max}$ is the maximal WIMP velocity in the halo
evaluated in the Earth frame;
vii) $ \frac{d\sigma}{dE_{R}}(v,E_{R}) = \left( \frac{d \sigma}{dE_{R}} \right)_{SI}+
\left( \frac{d \sigma}{dE_{R}} \right)_{SD} $,
with $\left( \frac{d \sigma}{dE_{R}} \right)_{SI}$ 
spin independent (SI) contribution and
$\left( \frac{d \sigma}{dE_{R}} \right)_{SD}$ 
spin dependent (SD) contribution.

Finally, the expected differential counting rate as a function of the detected energy, 
$E_{det}$, for a real multiple-nuclei detector (as e.g. the NaI(Tl)) when taking into account
the channeling effect
can easily be derived by summing the eq. (\ref{eq:rt}) over the nuclei species
and accounting for the detector energy resolution:

\begin{equation}
\frac{dR_{NaI}^{(ch)}}{dE_{det}}(E_{det}) =
\int G(E_{det},E') \sum_{A=Na,I}
\frac{dR^{(ch)}_A}{dE'}
(E') dE' \; .
\label{eq:labelmul}
\end{equation}

\noindent The $G(E_{det},E')$ kernel generally has a gaussian behaviour.

\begin{figure} [ht]
\centering
\vspace{-0.6cm}
\includegraphics[width=6.0cm] {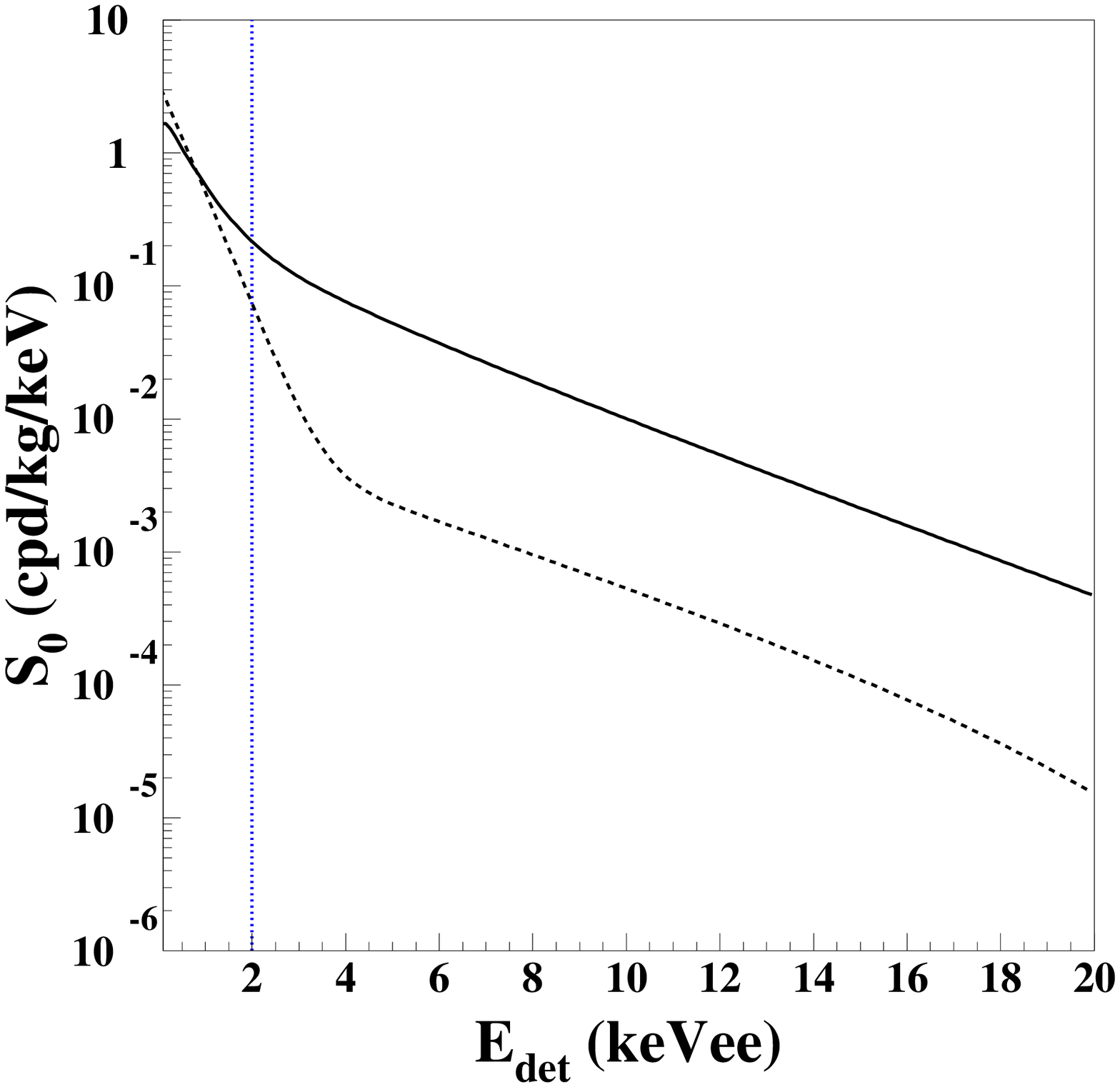}
\includegraphics[width=6.0cm] {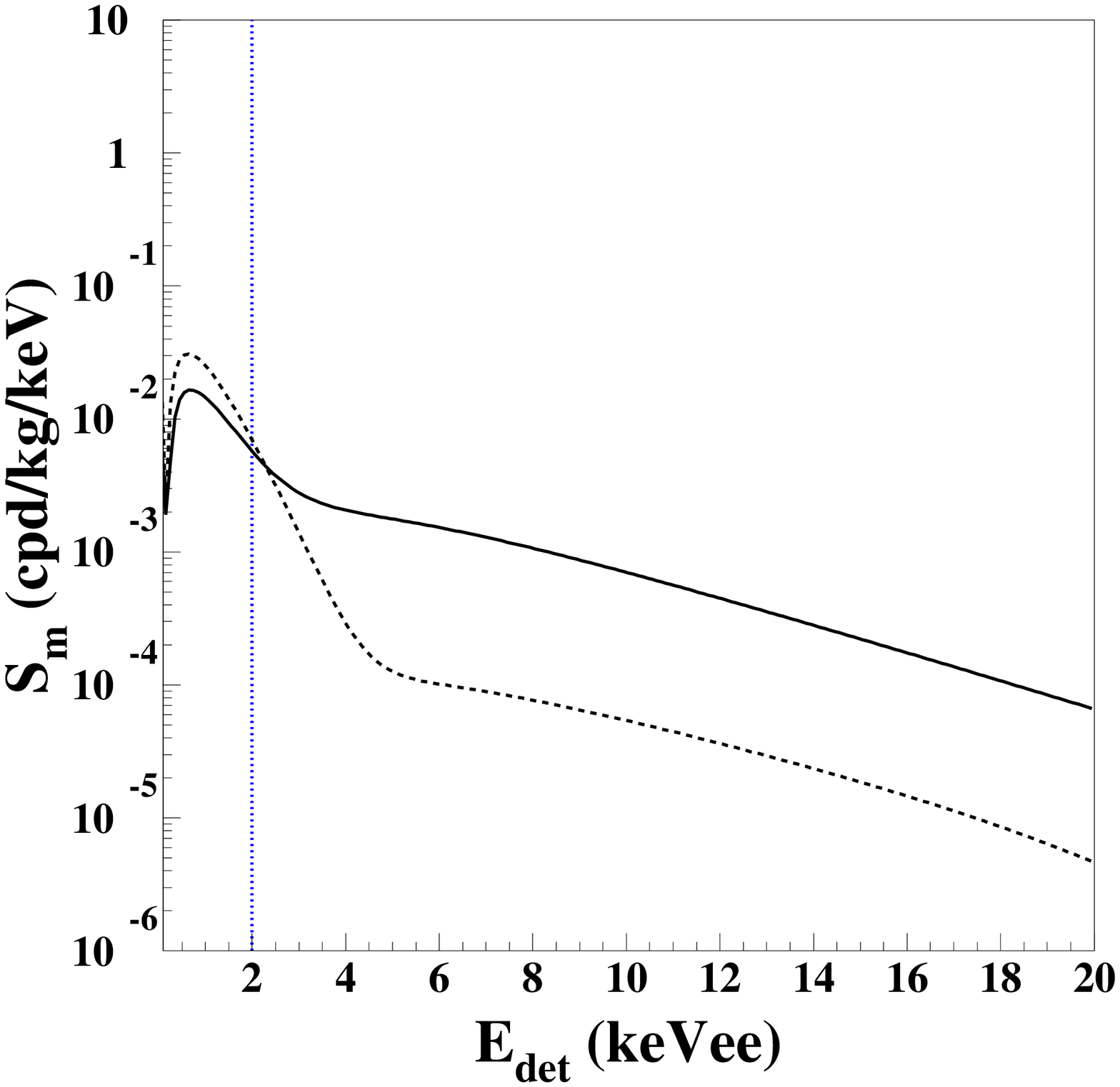}
\vspace{-0.4cm}
\caption{An example of the shapes of the expected energy distributions in NaI(Tl)
from Sodium and Iodine recoils induced by WIMP interactions
with (continuos line) and without (dashed line) including the channeling effect in the
crystal for the scenario given in the text. 
Left panel: behaviour of the unmodulated part of the expected signal, $S_0$.
Right panel: behaviour of the modulated part of the expected signal, $S_m$.
The vertical lines indicate the energy threshold of the DAMA/NaI experiment.}
\label{fg:spectra}
\end{figure}

Few examples of shapes of expected energy 
distributions with and without accounting for the channeling effect, calculated 
in the modeling given above, are shown in Fig. \ref{fg:spectra}.
For this template purpose -- accounting also for the experimental features of the detectors 
\cite{Nim98,Sist,RNC,ijmd} -- we have just adopted the following additional assumptions 
(among all the many possibilities):
i) WIMP mass of $m_W = 20$ GeV;
ii) WIMP with dominant Spin Independent coupling and with nuclear cross sections $\propto A^2$;
iii) point-like SI cross section $\sigma_{SI}=10^{-6}$ pb;
iv) an halo model NFW (identifier A5 in ref. \cite{halo}, local velocity $v_0 = 220$ km/s 
and halo density at the maximum value 0.74 GeV cm$^{-3}$ \cite{halo};
v) form factors and quenching factors of $^{23}$Na and $^{127}$I as in case A of ref. \cite{RNC}. 
These pictures point out the enhancement of the sensitivity due to the
channeling effect according to the given modeling.

\section{Examples of the possible impact on some corollary quests from the DAMA/NaI data}

The accounting of the channeling effect in corollary quests for WIMPs
as Dark Matter candidate particles can be investigated by exploiting  
the expected energy distribution, derived above, to some of the previous analyses 
on the DAMA/NaI annual modulation data
in terms of WIMP-nucleus elastic scattering. For this purpose, the same scaling laws 
and astrophysical, nuclear and
particles physics frameworks of refs. \cite{RNC,ijmd} are adopted. In addition, as already mentioned, 
for simplicity just to point out the impact of the channeling effect,
the possible SagDEG contribution to the galactic halo and the presence of the existing Migdal effect
-- whose effects we discussed in refs. \cite{epj06} and \cite{ijma2}, respectively --
will not be included here.

The results for each kind of interaction
are presented in terms of allowed volumes/regions,
obtained as superposition of the configurations corresponding
to likelihood function values {\it distant} more than $4\sigma$ from
the null hypothesis (absence of modulation) in each one of the several
(but still a very limited number) of the considered model frameworks.
This allows us to account -- at some extent -- for at least some of the 
existing theoretical and experimental uncertainties  
(see e.g. in  ref. \cite{RNC,ijmd,ijma,epj06,ijma2} and in the related astrophysics, nuclear and particle
physics literature). Here only the low mass volumes/regions, where the channeling effect is dominant, are depicted.

\vspace{0.3cm}

Since the $^{23}$Na and $^{127}$I are fully sensitive both to SI and to SD
interactions, the most general case is defined in a four-dimensional space
($m_W$, $\xi\sigma_{SI}$, $\xi\sigma_{SD}$, $\theta$), where: i) 
$\sigma_{SI}$ is the point-like SI WIMP-nucleon cross section and 
$\sigma_{SD}$ is the point-like SD WIMP-nucleon cross section, according to the definitions and 
scaling laws considered in ref. \cite{RNC}; ii) $tg\theta$ is the ratio between
the effective coupling strengths to neutron and proton for the 
SD couplings ($\theta$ can vary between 0 and $\pi$) \cite{RNC}.

\begin{figure} [!hb]
\centering
\includegraphics[width=5.5cm] {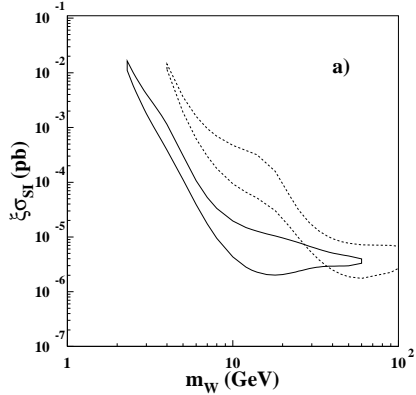}
\vspace{-0.4cm}
\caption{An example of the effect of the channeling, modelled as in the text, 
on a DAMA/NaI allowed region for purely SI coupling WIMPs in the given scenario (see text).
The region encloses configurations corresponding
to likelihood function values {\it distant} more than $4\sigma$ from
the null hypothesis (absence of modulation).
This example has been evaluated according to the assumptions given in the text.
In particular, an halo model Evans' logarithmic with $R_c=5$ kpc 
(identifier A1 in ref. \cite{halo}) has been considered 
for a $v_0$ value of 170 km/s and halo density at the corresponding maximum value \cite{halo};
the form factors parameters and the quenching factors 
of $^{23}$Na and $^{127}$I are as in case A of ref. \cite{RNC}. 
The solid (dashed) curves delimitate the allowed regions when the channeling effect is (not) included.
For simplicity just to point out the impact of the channeling effect,
the possible SagDEG contribution to the galactic halo and the presence of the existing Migdal effect
-- whose effects we discussed in refs. \cite{epj06} and \cite{ijma2}, respectively --
are not included here. Moreover, the same considerations reported in ref. \cite{RNC} still hold.}
\label{fg:reg1}
\end{figure}

Preliminarily, here to offer an example of the impact of accounting for the channeling effect as given in the text,
Fig. \ref{fg:reg1} shows a comparison for allowed slices corresponding to purely SI coupled WIMPs in some particular given 
scenario. 
This example has been evaluated for an halo model Evans' logarithmic with $R_c=5$ kpc 
(identifier A1 in ref. \cite{halo}) for a $v_0$ value of 170 km/s 
and halo density at the corresponding maximum value \cite{halo};
the form factors parameters and the quenching factors 
of $^{23}$Na and $^{127}$I are as in case A of ref. \cite{RNC}. 
The solid (dashed) curves delimitate the allowed regions in the given scenario 
when the channeling effect is (not) included.
As it can be seen, for WIMP masses in the few-20 GeV region the allowed SI region 
when including the channeling effect is lower than one order of magnitude in cross section. 

\begin{figure} [!t]
\centering
\vspace{-1.4cm}
\includegraphics[width=6.5cm] {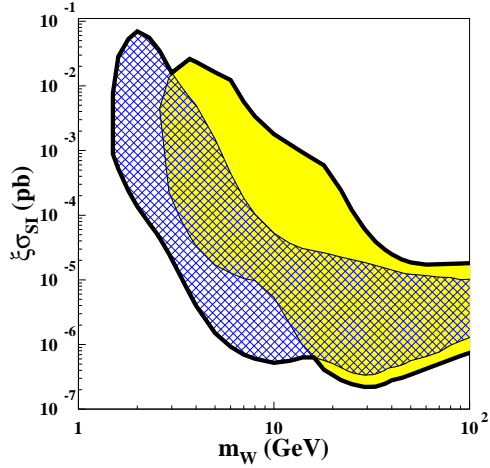}
\vspace{-0.4cm}
\caption{Region allowed in the ($\xi\sigma_{SI},m_W$) plane in the same model frameworks 
of ref. [10,11] for pure SI coupling; 
just the low mass part of interest for the channeling effect 
is focused here. The dotted region is obtained in absence of channeling effect [10,11], while 
the dashed one is obtained 
when accounting for it as described in the text.  
The dark line marks the overal external contour. It is worth to note that the inclusion of other 
contributions and/or of other uncertainties on parameters and models,
such as e.g. the possible SagDEG contribution [12] and the Migdal effect [13] or more favourable form factors, 
different scaling laws,  
etc., would further extend the region and increases the sets of the best fit values.
For completeness and more see also [10-14].
The same considerations reported in the caption of Fig. \ref{fg:reg1} hold.}
\label{fg:reg_si}
\vspace{-0.4cm}
\end{figure}

\begin{figure} [!hb]
\centering
\vspace{-1.2cm}
\includegraphics[width=9.0cm] {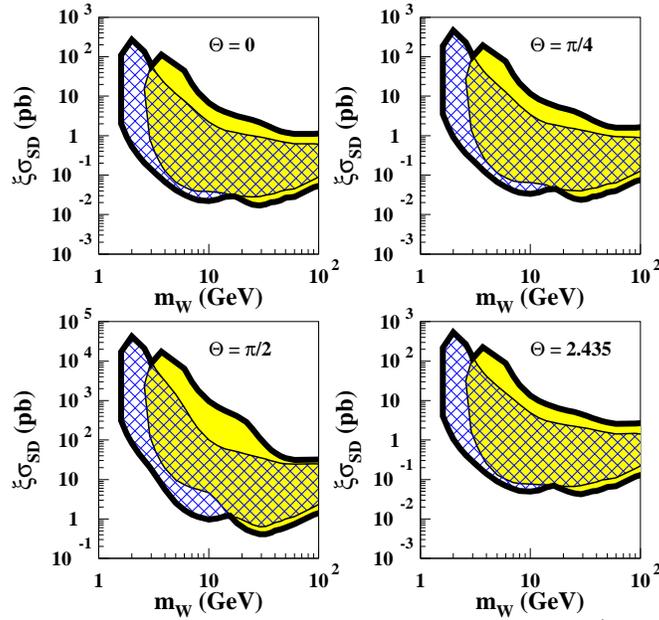}
\vspace{-0.6cm}
\caption{Examples of slices of the 3-dimensional allowed volume 
($\xi\sigma_{SD},m_W,\theta$)
in the same model frameworks of ref. [10,11] for pure SD coupling; just the low mass part of interest 
for the channeling effect is focused here. 
Analogous comments and remarks as those in the captions of Figs. \ref{fg:reg1} and \ref{fg:reg_si} hold.}
\label{fg:reg_sd}
\vspace{-0.5cm}
\end{figure}

The subcase of purely SI coupled WIMPs for the scenarios of ref. \cite{RNC,ijmd} is shown in Fig. \ref{fg:reg_si},
while in Fig. \ref{fg:reg_sd} four
slices of the 3-dimensional allowed volume ($m_W$, $\xi\sigma_{SD}$, $\theta$) for the purely
SD case are given as example; the low mass region of interest for the effect is just focused here.

\begin{figure} [!ht]
\centering
\vspace{-0.5cm}
\includegraphics[width=10.0cm] {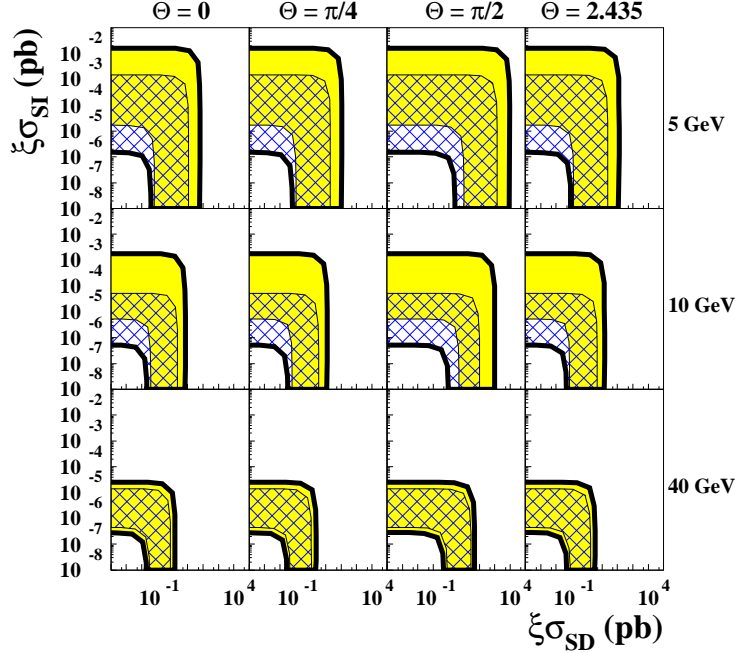}
\vspace{-3.0cm}
\caption{Examples of slices of the 4-dimensional allowed volume 
($\xi\sigma_{SI},\xi\sigma_{SD},m_W,\theta$) in the model frameworks
considered in ref. [10,11] for mixed SI\&SD coupling; just the low mass part of interest 
for the channeling effect is focused here.
Analogous comments and remarks as those in the captions of Figs. \ref{fg:reg1} and \ref{fg:reg_si} hold.}
\label{fg:reg_sisd}
\end{figure}

Finally, in the general case of mixed SI\&SD coupling 
one gets, as mentioned above, a 4-dimensional allowed volume ($\xi\sigma_{SI},\xi\sigma_{SD},m_W,\theta$).
New allowed volume at the given C.L. is present in the GeV region when accounting for the channeling effect.
Fig.\ref{fg:reg_sisd} shows few slices of such a volume as examples.

Note that general comments, extensions, etc.  already 
discussed in ref. \cite{RNC,ijmd,ijma,epj06,ijma2}
still hold.

\section{Conclusions}

In this paper the channeling effect of recoiling nuclei induced by WIMP and WIMP-like elastic
scatterings in NaI(Tl) crystals has been discussed. Its possible effect in a reasonably 
cautious modeling has been presented as applied to some given simplified scenarios 
in corollary quests for the 
candidate particle for the DAMA/NaI model independent evidence. 
This further shows the role of the existing uncertainties and of the correct 
description and modeling of all the involved processes as well as
their possible impact in the investigation of the candidate 
particle. Some of them have already been addressed at some extent, such as the
halo modeling \cite{halo,RNC,ijmd}, the possible presence of 
non-thermalized components in the halo (e.g. caustics \cite{siki} or
SagDEG \cite{epj06} contributions), the accounting for the electromagnetic 
contribution to the WIMP (or WIMP-like) expected energy distribution \cite{ijma2}, 
candidates other than WIMPs (e.g. \cite{ijma} and in literature), etc.. 

Obviously, many other arguments can be addressed as well both on DM candidate particles
and on astrophysical, nuclear and particle physics aspects;
for more see \cite{RNC,ijmd,ijma,epj06,ijma2} and in literature.
In particular, we remind that different astrophysical, nuclear and particle 
Physics scenarios as well as the experimental and theoretical associated uncertainties 
leave very large space also e.g. for significantly lower cross sections and larger masses.

\end{document}